\documentclass[onecolumn]{emulateapj}

\usepackage{epsfig}

\shorttitle{Alcohols \& Thiols}
\shortauthors{Gorai et al.}

\begin{document}

\title{  Search for Interstellar monohydric Thiols}

\author{Prasanta Gorai\altaffilmark{1}, Ankan Das\altaffilmark{1}, Amaresh Das\altaffilmark{2,1}, 
Bhalamurugan Sivaraman\altaffilmark{3}, Emmanuel E. Etim\altaffilmark{4,5}, 
Sandip K. Chakrabarti\altaffilmark{6,1}}
\affil{$^1$Indian Centre for Space Physics, 43 Chalantika, Garia Station Rd., Kolkata, 700084, India}
\affil{$^2$ Ramakrishna Mission Residential College, Narendrapur, Kolkata 700103, West Bengal, India}
\affil{$^3$ Atomic Molecular and Optical Physics Division, Physical Research Laboratory, Ahmedabad, 380009, India}
\affil{$^4$ Indian Institute of Science Bangalore, India-560012}
\affil{$^5$ Department of Chemical Sciences, Federal University Wukari, Nigeria}
\affil{$^6$ S.N. Bose National Centre for Basic Sciences, Salt Lake, Kolkata, 700106, India}

\email{$^1$ankan.das@gmail.com}

\begin{abstract}
It has been pointed out by various astronomers that very interesting relationship
exists between interstellar alcohols and the corresponding thiols (sulfur analogue of alcohols)
as far as the spectroscopic properties and chemical abundances are concerned. Monohydric alcohols such as 
methanol and ethanol are widely observed and 1-propanol is recently claimed to have been seen in Orion KL.
Among the monohydric thiols, methanethiol (chemical analogue of methanol), 
has been firmly detected in Orion KL and Sgr B2(N2) and ethanethiol 
(chemical analogue of ethanol) has been claimed to be observed in Sgr B2(N2) though the confirmation of 
this detection is yet to come. It is very likely that higher order thiols could be 
observed in these regions. In this paper, we study the formation of monohydric
alcohols and their thiol analogues. Based on our quantum chemical calculation and chemical
modeling, we find that `Tg' conformer of 1-propanethiol is a good
candidate of astronomical interest. We present various spectroscopically relevant 
parameters of this molecule to assist its future detection in the Interstellar medium (ISM).
\end{abstract}


\section{Introduction}
Starting from the detection of first carbon containing molecule, methylidyne radical (CH) 
in 1937 \citep{swin37}, almost $200$ molecules including neutrals, radicals and
ions have been observed in the interstellar medium or circumstellar shells and 
almost $60$ molecules have been observed in comets. 
A mismatch between the cosmic abundance of sulfur and observed abundances of S-bearing 
species is well known \citep{palu97}. Particularly around the dense cloud regions, 
this inequality is severe \citep{tief94,palu97}. Around the diffuse cloud and highly 
ionized regions, sulfur related species roughly resemble the cosmic abundance $\sim10^{-5}$ 
\citep{sava96,howk06}.
Earlier, \cite{mill90,jans95} suggested that S, SO, CS and $\rm{H_2S}$
may explain the missing sulfur problem though our knowledge about the CS related species is very limited.
Recently \cite{mull15}
suggested that at $400$ K more than $50$\% of the sulfur budget is shared by 
CS and $\rm{H_2CS}$ and remainder resides in the form of SO and SO$_2$ for hot source Sgr B2(N).
Several experiments were carried out to propose the abundant
S-bearing species on interstellar grains. Outcome of these experiments proposed that OCS
\citep{garo10}, $\rm{CS_2}$ \citep{ferr08}, hydrated
sulfuric acid \citep{scap03} would act as a sink for the interstellar sulfur.
Till date, only two sulfur related 
molecules (OCS and $\rm{SO_2}$) had been detected on grain surface with full confidence thus the 
exact reservoir of sulfur is yet to be known with certainty
\citep{wood15,palu95,boog97}. 

Among the monohydric alcohols, methanol ($\rm{CH_3OH}$) is the simplest alcohol which is 
widely observed both in gas and solid phases \citep{tiel87a} of the ISM. 
Major portion of the interstellar grain mantle is
found to be covered with methanol \citep{gibb04,das08a,das10,das11,das16}. 
Gas phase abundance of methanol relative to ${\rm H_2}$ is found 
to be in the range of $10^{-9}$ in cold dark clouds to $10^{-6}$ in hot molecular 
cores \citep{char95}. The presence of Ethanol ($\rm{C_2H_5OH}$) (second alcohol in 
this homologous series) is observed in  star-forming regions in the range of $10^{-8}-10^{-6}$ 
\citep{mill88,turn91}. Propanol is the next alcohol in the series of 
monohydric alcohols which may belong in two different forms: normal(n)-propanol (${\rm CH_3CH_2CH_2OH}$) 
and 2-propanol (${\rm CH_3CHOHCH_3}$). Recently n-propanol (1-propanol) was claimed to be 
detected towards Orion KL by \cite{terc15} with a column density 
of $\leqslant (1.0\pm 0.2 \times 10^{15})$ cm$^{-2}$ whereas the presence of 2-propanol is yet to 
be verified.

It is now confirmed that methanol and ethanol are mainly produced on dust grains during 
the cold phase and evaporate from warm dust grains in latter
stages of evolution. Following this trend, even higher order alcohols would be produced 
on interstellar grains. 
In case of thiols, sulfur takes the place of oxygen in the hydroxyl group of an alcohol. 
Similar to their alcohol analogues, these thiols are mainly produced on the grain surface
and are evaporated in suitable time.
Tentative  detection of Methanethiol ($\mathrm{CH_3SH}$) in Sgr B2 had 
been done by \cite{turn77}. Later this claim had been confirmed by \cite{link79} and 
showed that $\mathrm{CH_3SH/CH_3OH}$ ratio is close to the cosmic S/O ratio. Recently,
\cite{maju16} detected $\rm{CH_3SH}$ in IRAS 16293-2422 and
\cite{kole14} reported the detection of  $\mathrm{C_2H_5SH}$ in hot core 
Orion KL. But very recent observation by \cite{mull15} suggested that the detection 
of $\mathrm{C_2H_5SH}$ in Orion KL is uncertain. Presence of higher order thiols are yet to be 
seen.

In this paper, we discuss the formation of monohydric alcohols and their 
thiol analogues. First, we identify the most stable
conformer of alcohols and their thiols. Then we develop a chemical network to study
the formation of all these species. From the outcome of our chemical modeling, 
most probable new candidate for the astronomical detection
is found out. Moreover, a detailed spectroscopic study is carried out to set a
guideline for observing this species in near future.
This paper is organized as follows: Section 2 describes the quantum chemical calculation
to find out the most stable conformers; Section 3
contains chemical modeling; Section 4
contains the information regarding the spectroscopic parameters for the detection of the
next probable candidate; Finally in Section 5, we make concluding remarks.

 \section{Search for most stable conformational isomers}
 All the quantum chemical calculations reported here are calculated by using Gaussian 09 
program \citep{fris09,fore96}. 
Each optimized structure is verified by avoiding 
the imaginary frequency. With the advance of quantum chemical calculation a proper 
choice of Method and basis sets are required to compute the molecular properties.

\begin{figure}
\centering
\includegraphics[width=8.3cm, height=5cm]{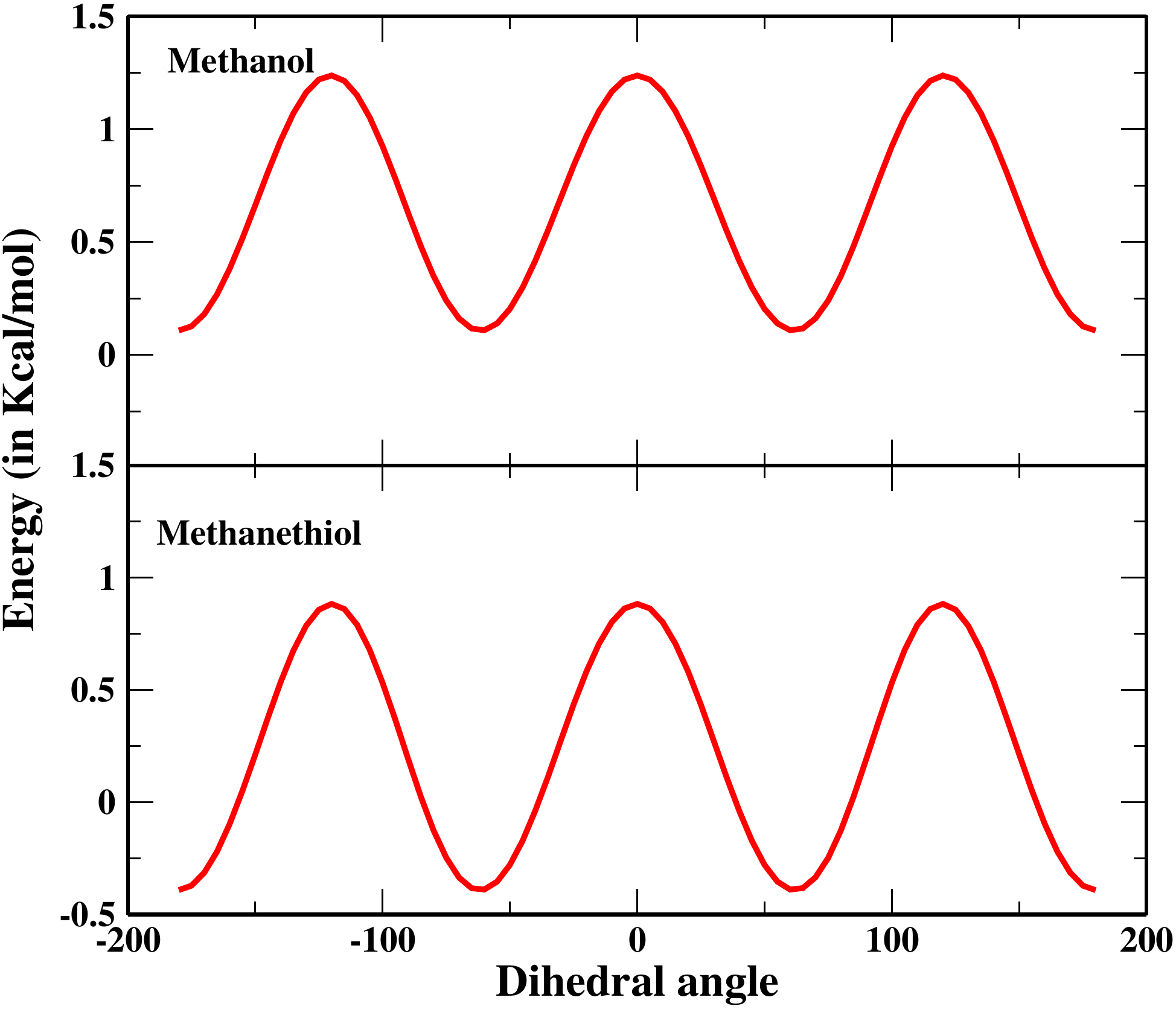}
\includegraphics[width=7cm, height=5cm]{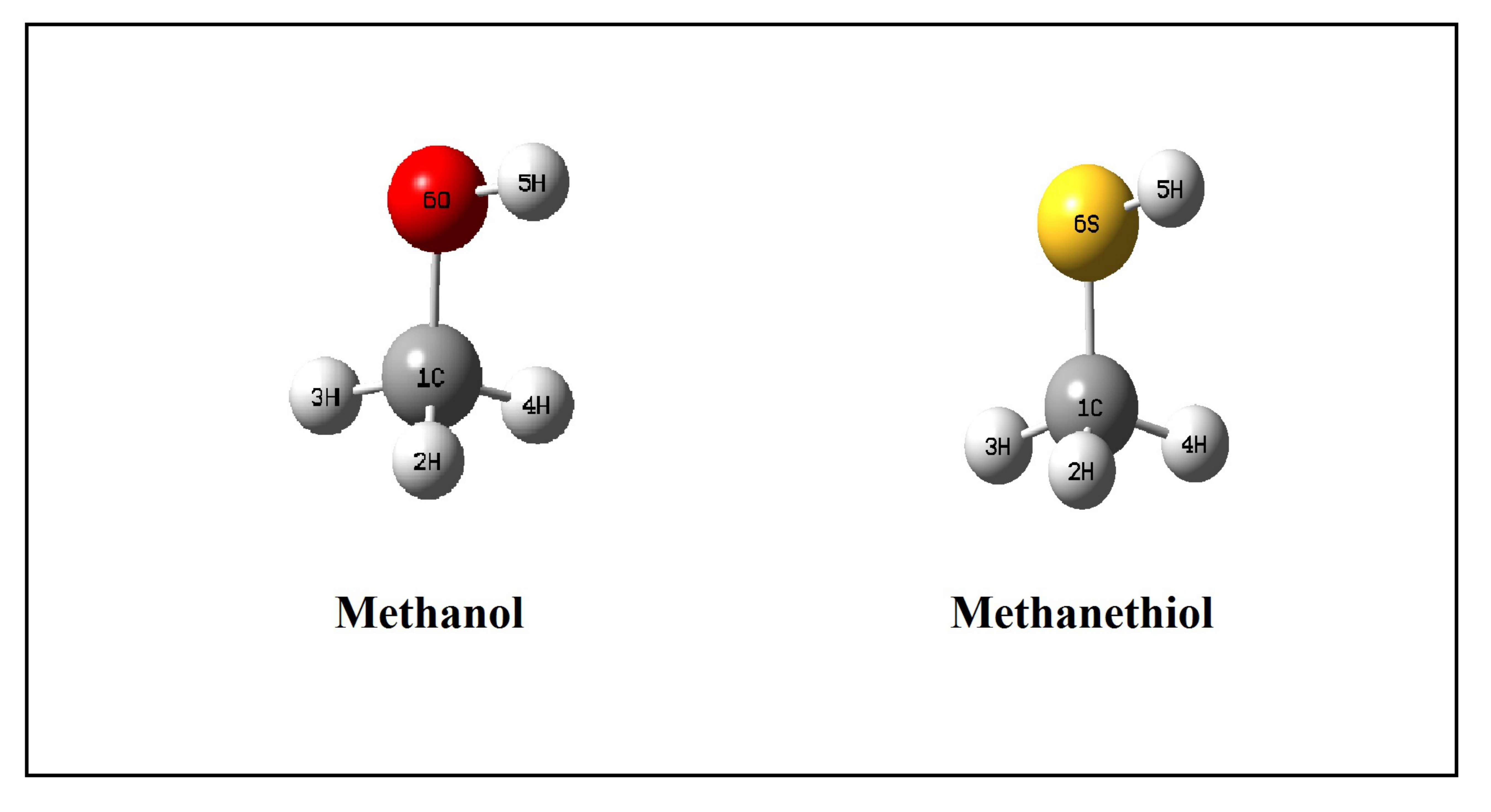}
\vskip 0.4cm
\includegraphics[width=8cm, height=5cm]{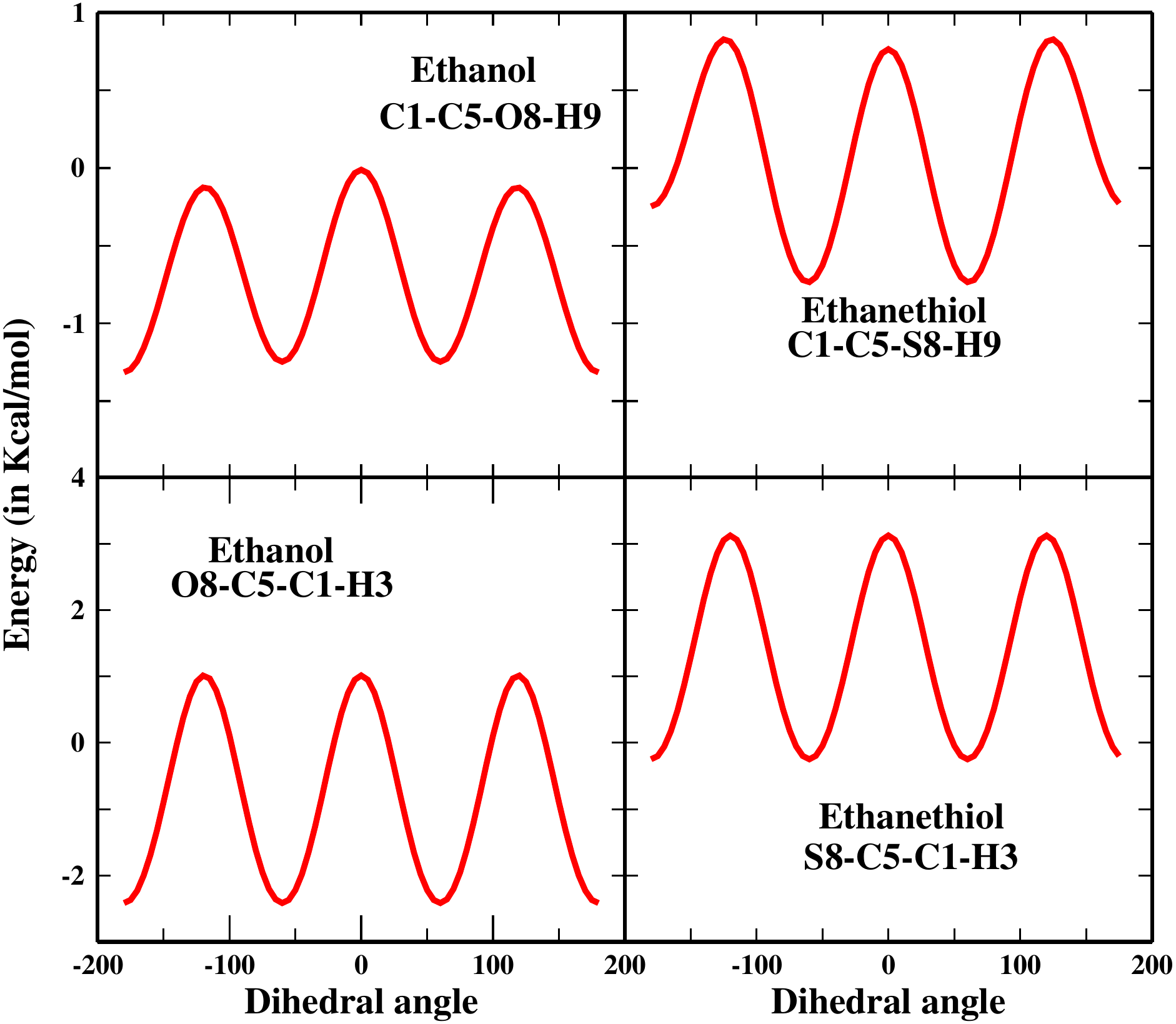}
\includegraphics[width=7cm, height=5cm]{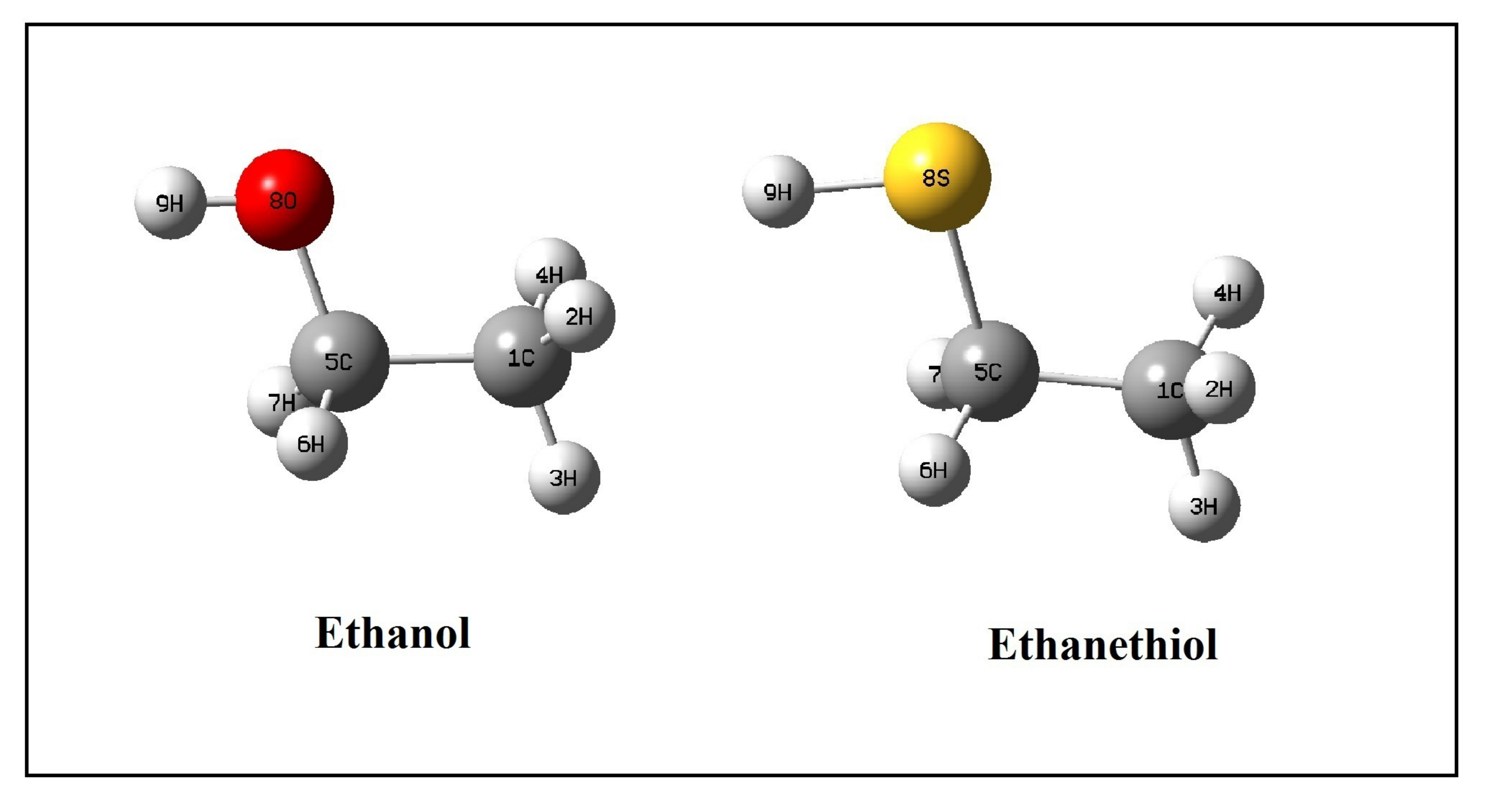}
\vskip 0.4cm
\includegraphics[width=8cm, height=5cm]{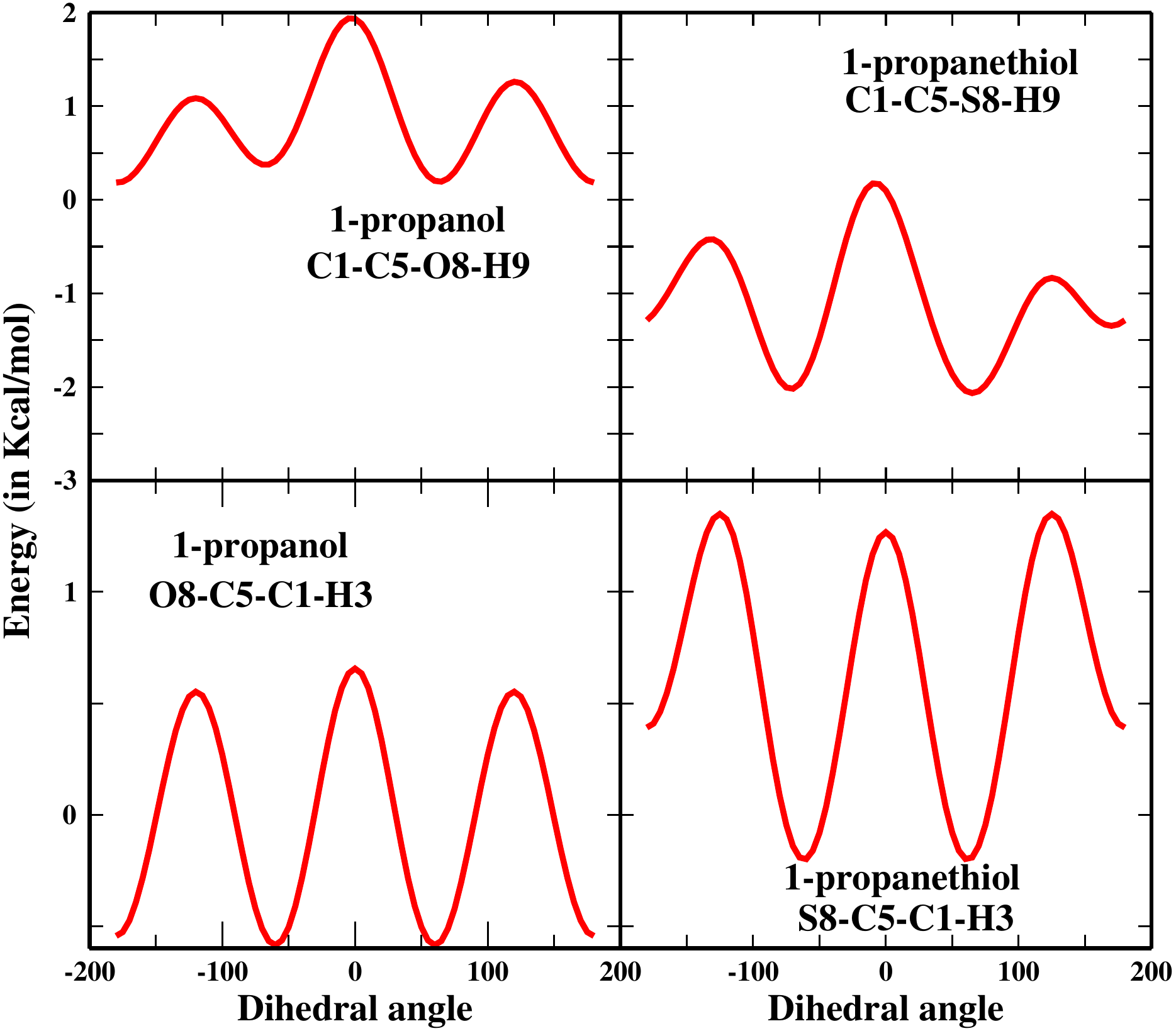}
\includegraphics[width=7cm, height=5cm]{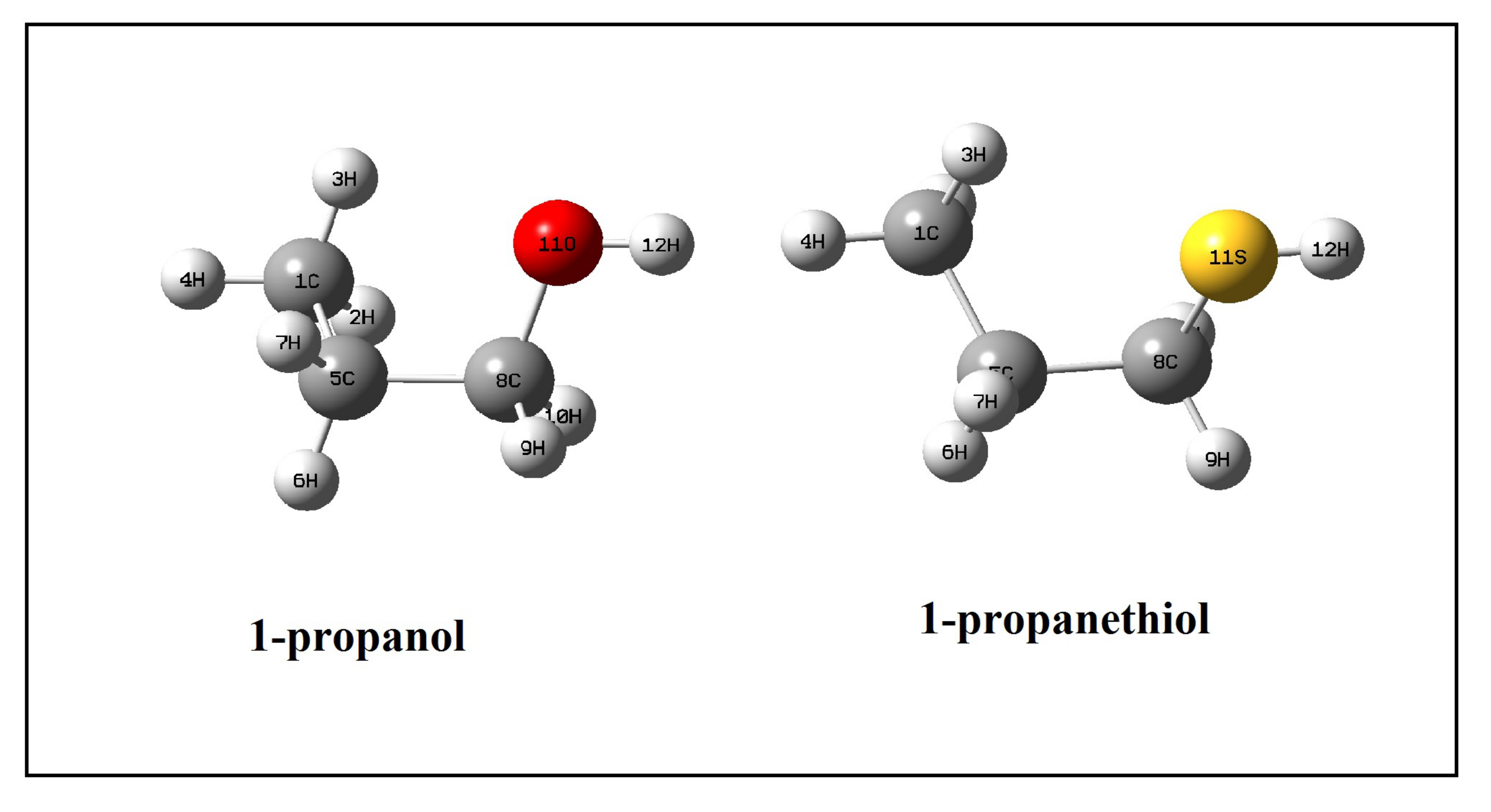}
\vskip 0.4cm
\includegraphics[width=8.3cm, height=5cm]{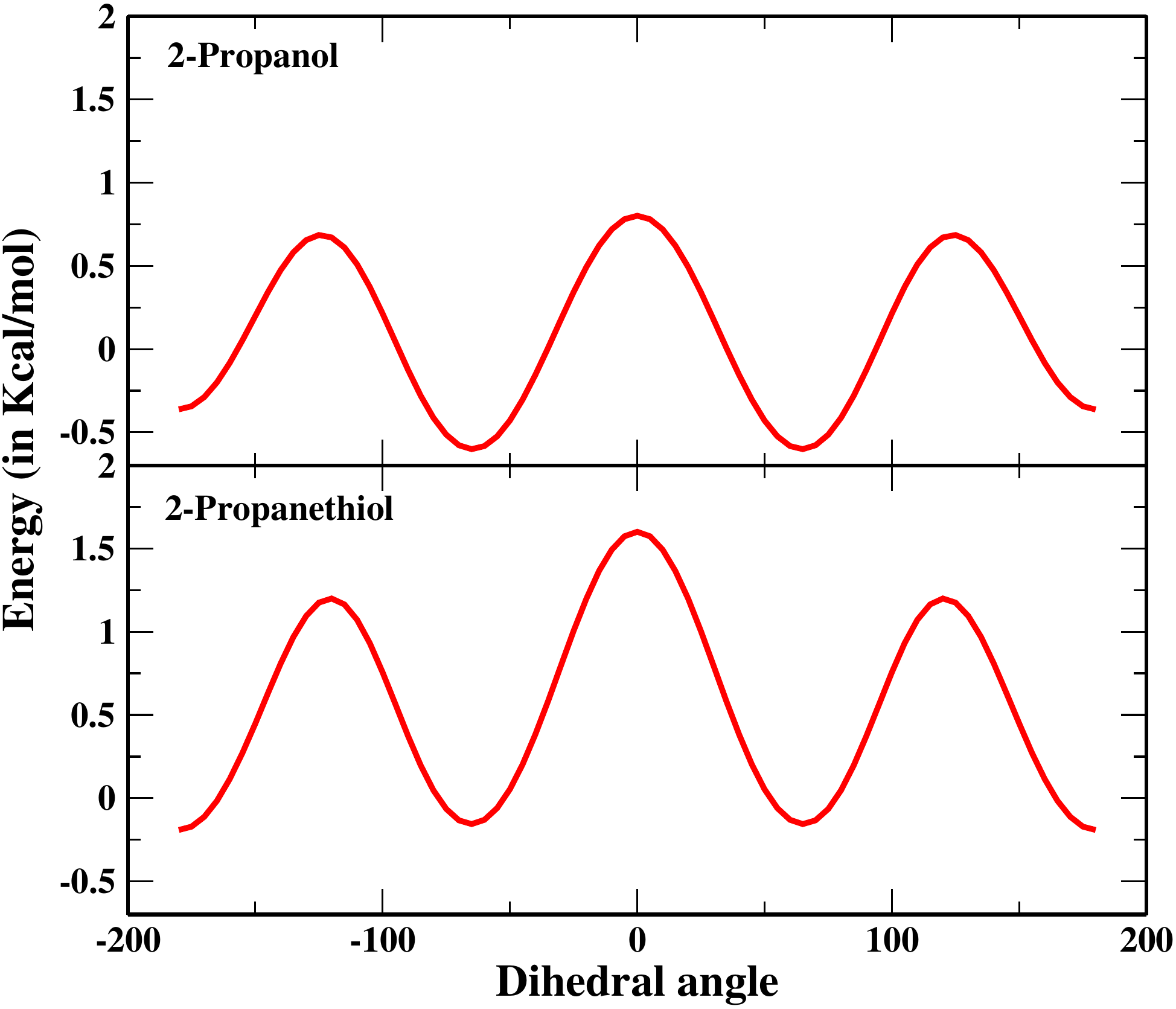}
\includegraphics[width=7cm, height=5cm]{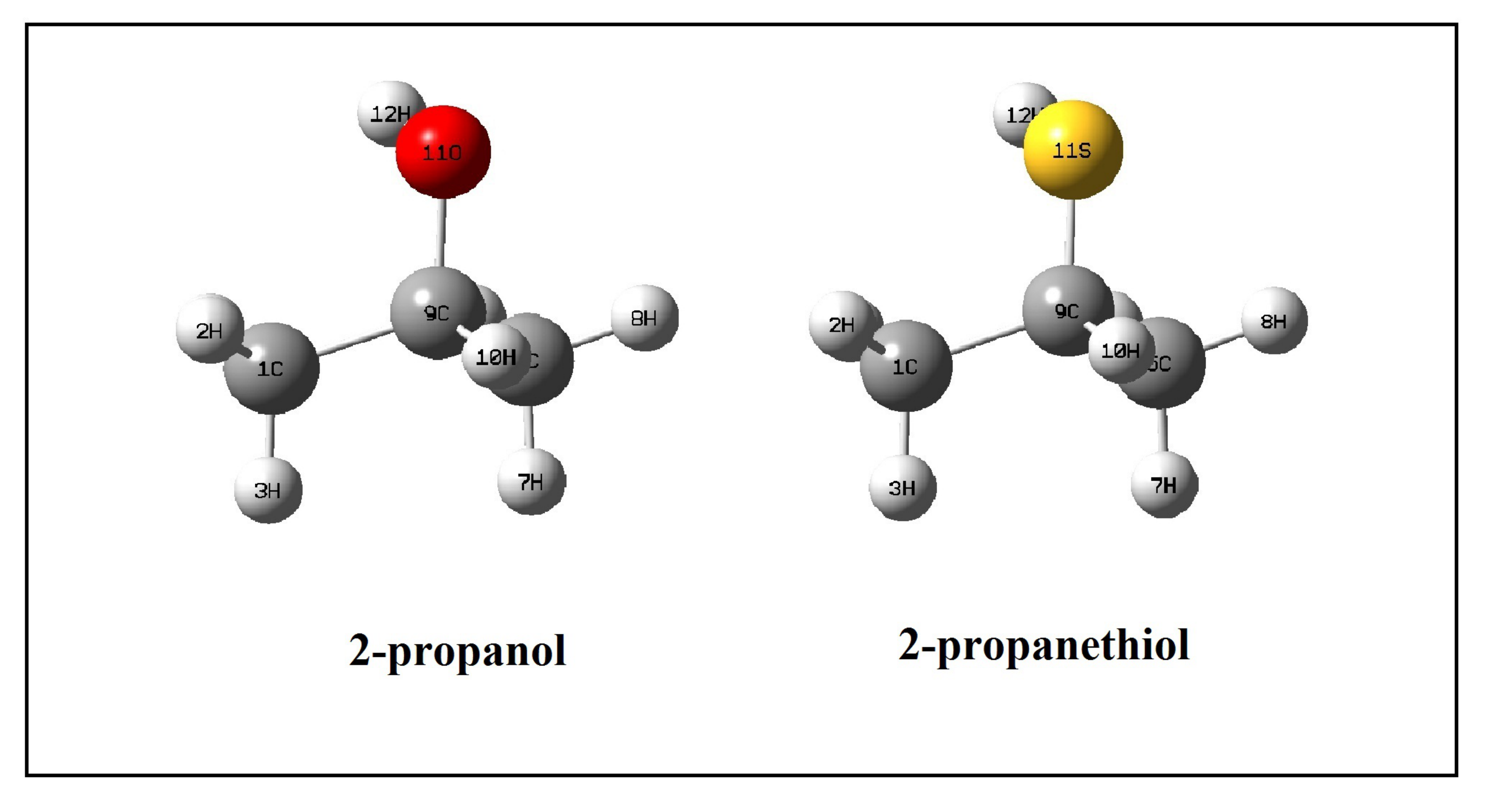}
\caption{\scriptsize Relaxed potential energy surface scan  of dihedral angle of monohydric alcohols and their thiol analogues using MP2/cc-pVTZ level of theory.}
\vskip 1cm
\label{fig-1}
\end{figure}

According to IUPAC definition of a conformer, conformational isomerism is a form 
of stereo isomerism in which the isomers
can be inter-converted exclusively by rotations about formally single bonds. It is expected that the most
stable conformer would be the most probable candidate for the astronomical detection. 
In this attempt, here, before constructing our chemical model, we search for the various 
conformers of the monohydric alcohols and their thiols through relaxed potential energy 
surface (PES) scan of dihedral angles. PES scan results are displayed in Fig. 1 and relative 
energies of the conformers
are pointed out in Table 1. 2-propanol/2-propanethiol is the
structural isomer of 1-propanol/1-propanethiol, for the
sake of completeness, we discuss about their conformers as well. 
{For all these  calculation, we use M\o ller-Plesset perturbation theory (MP2) 
with the Peterson and Dunning's correlation consistent basis set (cc-pVTZ) \citep{pete02} 
of Gaussian 09 software.} Results of the PES scans are the following.

\subsection{Methanol \& methanethiol}
Methanol and methanethiol both show internal rotation of $\rm{CH_3}$ group. From relaxed
potential energy surface scan ($-180^{\circ}$ to $+180^{\circ}$ of dihedral angle $\angle\rm{(H_3, C_1, O_6, H_5)}$),
we have shown that methanol and methanethiol both exist in the most stable state at $\pm$ $180^{\circ}$ 
of the dihedral angle on PES. On the other hand, there exist three different maxima for both methanol
and methanethiol at $\pm$ $120^{\circ}$ and at $0^{\circ}$. Relative energies are pointed out in Table 1.

\begin{table}
\caption{Relative energies of various conformers of alcohols and their thiol analogues}
\vbox
{\centering \tiny
\begin{tabular}{|c|c|c|}
\hline
{\bf Species}&{\bf Conformer}&{\bf $\Delta E$ in cm$^{-1}$ (Kcal/mol)}\\
\hline \hline
&HCOH $\pm$ $180^{\circ}$& 0 (0) \\
&HCOH $\pm$ $60^{\circ}$& 0.4389 (0.0013)\\
{\bf Methanol}& HCOH $120^{\circ}$&395 (1.13)\\
&HCOH $0^{\circ}$&395 (1.13)\\
&&\\
\hline
&HCSH $\pm$ $180^{\circ}$&0 (0)\\
&HCSH $\pm$ $60^{\circ}$&0.877 (0.0025)\\
{\bf Methanethiol}&HCSH $\pm$ $120^{\circ}$&445 (1.27)\\
&HCSH $0^{\circ}$&445 (1.27)\\
&&\\
\hline
&$trans$ &0 (0)\\
&$gauche$ &23 (0.066)\\
{\bf Ethanol}&$eclipsed$&415 (1.19)\\
&$cis$&456 (1.30)\\
&&\\
\hline
&$gauche$ CCSH &0 (0)\\
&$trans$ CCSH &170.0 (0.49)\\
{\bf Ethanethiol}&$eclipsed$ &524.0 (1.50)\\
&$ cis$ &541 (1.55)\\
&&\\
\hline
&Gt& 0 (0)\\
&Gg&67 (0.19)\\
{\bf 1-propanol}&$Gg$ $^{'}$&3.5 (0.01)\\
&$Tt$&95 (0.27)\\
&$Tg$&80 (0.23)\\
\hline
&$Tg$&0 (0)\\
&$Tt$&216 (0.62)\\
{\bf 1-propanethiol}&Gt&318 (0.91)\\
&$Gg$ $^{'}$&63 (0.18)\\
&$Gg$&46 (0.13)\\
\hline
&$gauche$&0 (0)\\
{\bf 2-propanol}&$trans$&83.6 (0.24)\\
\hline
&$trans$&0 (0)\\
{\bf2-propanethiol}&gauche&12 (0.034)\\
\hline
\end{tabular}}
\vskip 1cm
\end{table}

\subsection{Ethanol \& ethanethiol}

In case of ethanol and ethanethiol, both have two types of internal rotation: around OH/SH group and CH$_3$
group. Due to internal rotation of the OH/SH group of ethanol/ethanethiol, it may
exist in four different forms. For ethanol, trans/anti conformation (dihedral angle 
$\angle{\rm (C_1, C_5, O_8, H_9)}$  =$180^{\circ}$) is the  minimum energy state on PES. The gauche
conformer (dihedral angle $\angle{\rm(C_1, C_5, O_8, H_9)}  = \pm 60^{\circ}$) is situated
slightly upper on the PES. The relative energy between the trans and gauche conformer
is $0.877$ cm$^{-1}$. When the dihedral angle is $\pm$ $120^{\circ}$,
conformers are called eclipsed. The relative
energy between trans and eclipsed conformer is $415$ cm$^{-1}$.
When the dihedral angle is $0^{\circ}$, it is called
cis conformer which is also situated at higher energy state  on the PES. The relative energy between 
trans and cis conformer is $456$ cm$^{-1}$ (1.3 Kcal/mol). Whereas for ethanethiol the most stable conformer is gauche in
which $\angle{\rm (C_1, C_5, S_8, H_9)}$ dihedral angle is at $\pm$ $60^{\circ}$. Relative energy between gauche and
trans conformer is $170$ cm$^{-1}$ (0.49 Kcal/mol), relative energy between gauche and eclipsed conformer 
is $524$ cm$^{-1}$ (1.50 Kcal/mol) and the relative energy between gauche and cis conformer is 
$541$ cm$^{-1}$ (1.55 Kcal/mol). 
Due to ${\rm CH_3}$ rotation, there would also be some changes in the energy between the conformers
but all are higher as compared to the most stable one.

\subsection{1-propanol and 1-propanethiol}

\cite{maed06} discussed five possible conformational isomers of 1-propanol
originating from alternate structure of the central $\rm{(C_1, C_5, O_8, H_9)}$ and $\rm{(O_8, C_5, C_1, H_3)}$
skeletal chains. These five forms are $Trans-trans\ (Tt),\ Trans-gauche \ (Tg),\ 
Gauche-trans\ (Gt),\ Gauche-gauche \ (Gg)$ and 
$Gauche-gauche^{'}$ 
$(Gg^{'})$. Based on the relative energies of these conformers, \cite{abdu87} find out that
$Tg$ conformer has the lowest energy. However \cite{lott84,kisi10} predicted that $Gt$ conformer
has the lowest energy. Our quantum chemical study also finds that $Gt$ conformer has the lowest energy. 
As in earlier studies, in terms of their relative energies, we found that 1-propanol conformer follows
the sequence of $Gt,\ Gg^{'}, \ Gg, \ Tg, \ Tt$ in the ascending order of relative energies.

Similar studies have been carried out for 1-propanethiol. Since 1-propanethiol is yet to be
detected in the ISM, perspective views of its five possible staggered conformers, originating
from the combination of $trans$ and $gauche$ configuration is shown in Fig. 2. 
Our calculation reveals that $Tg$ configuration has the lowest energies. Relative energies 
of all the conformers with respect to the $Tg$ conformer are shown in Table 1.

\subsection{2-propanol and 2-propanethiol}
Depending on the rotation of the dihedral angle $\angle {\rm(H_{10}, C_9, O_{11}, H_{12})}$/$\angle {\rm (H_{10}, C_9, S_{11}, H_{12})}$, 
2-propanol/2-propanethiol may exist in two forms; $gauche$ and $trans$. $gauche$ conformer is found
to be stable for 2-propanol whereas, in case of 2-propanethiol, $trans$ conformer is
found to be stable.
Since 2-propanol/2-propanethiol is a
secondary alcohol/thiol, we are not considering this in our chemical modeling.

\begin{figure}
{\centering
  \includegraphics[height=8cm,width=8cm]{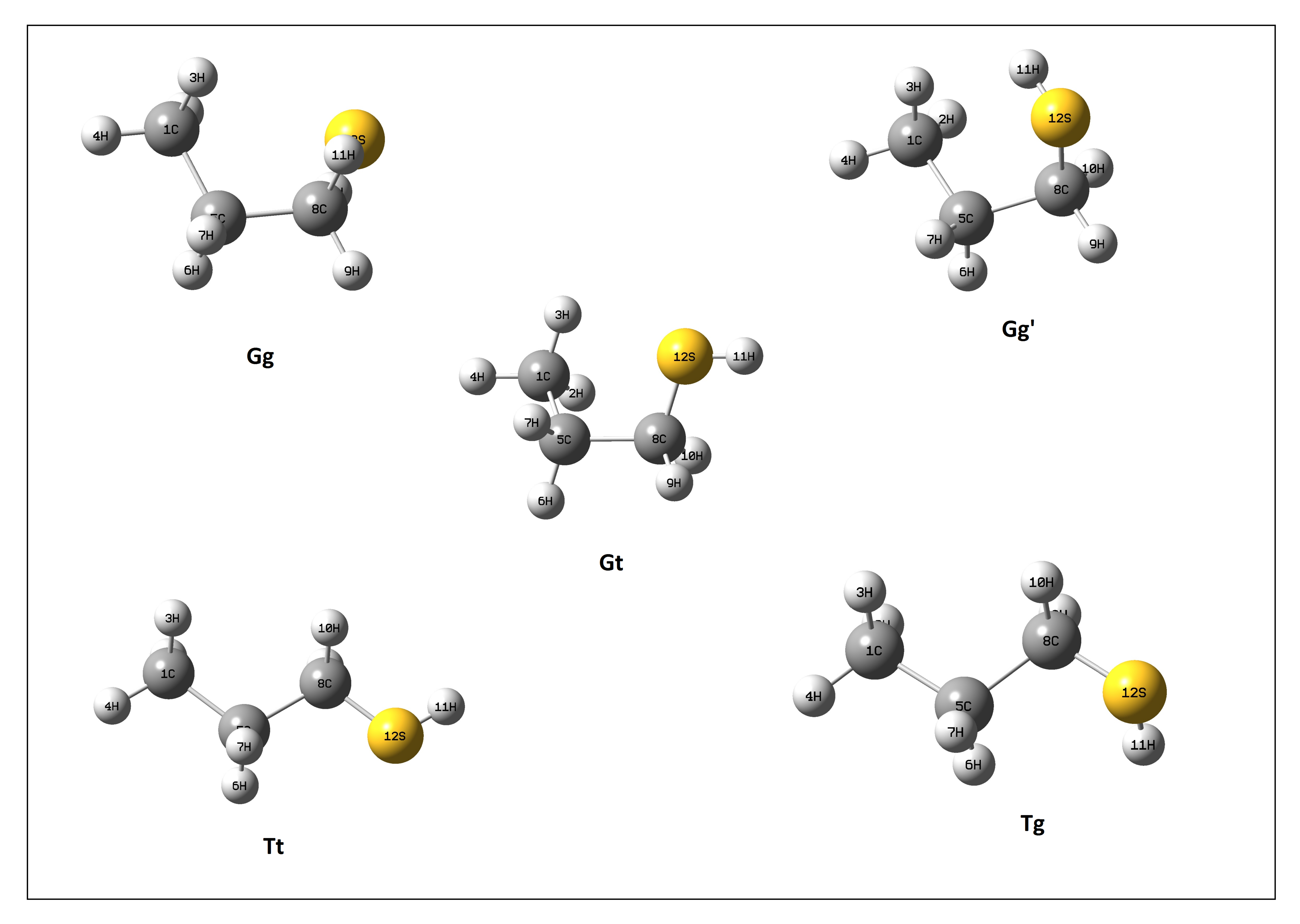}
\caption{\scriptsize Five possible conformers of 1-propanethiol.}}
\vskip 0.5cm
\end{figure}

\section{Chemical modeling}

\begin{table}
\vbox{\centering
{\tiny 
\caption{Ice phase production of alcohols and their corresponding thiols}
\begin{tabular}{|l|c|c|}
\hline
\hskip 2cm {\bf Reaction}&{\bf Energy Barrier (K)}&Reference\\
\hline
\hline
\hskip 2cm {\bf Methanol}&&\\
$\mathrm{H + CO  \rightarrow HCO}$ & 1000&H \\
$\mathrm{H + CO  \rightarrow HOC }$& 1000&H \\
$\mathrm{H + HCO \rightarrow H_2CO }$& 0&H \\
$\mathrm{H + HOC \rightarrow CHOH }$& 0&H \\
$\mathrm{CH + OH  \rightarrow CHOH }$& 0&H \\
$\mathrm{H + H_2CO \rightarrow HCO +H_2 }$& 1850&H \\
$\mathrm{H + CHOH \rightarrow CH_2OH }$& 0&H \\
$\mathrm{OH+ CH_2   \rightarrow CH_2OH }$& 0&H \\
$\mathrm{O + CH_3   \rightarrow CH_2OH }$& 0&H \\
$\mathrm{H + CH_2OH \rightarrow CH_3OH }$& 0&H \\
\hskip 2cm{\bf Methanethiol}&&\\
$\mathrm{H + CS  \rightarrow HCS}$ & 1000&H \\
$\mathrm{H + HCS  \rightarrow H_2CS}$ &    0&H \\
$\mathrm{H + H_2CS  \rightarrow CH_3S}$ &  1000&M \\
$\mathrm{S + CH_3   \rightarrow CH_3S}$ &     0 &  \\
$\mathrm{H + H_2CS  \rightarrow HCS + H2}$ &  1000&M \\
$\mathrm{H + H_2CS  \rightarrow CH_2SH}$ &  1000&M \\
$\mathrm{CH_2 + HS  \rightarrow CH_2SH}$ &    0&  \\
$\mathrm{H + CH_3S  \rightarrow CH_3SH}$ &     0&M \\
$\mathrm{H + CH_2SH  \rightarrow CH_3SH}$ &     & \\
\hline
\hskip 2cm{\bf Ethanol} &&\\
$\mathrm{C_2H_5 + OH  \rightarrow C_2H_5OH}$ &     0& H\\
$\mathrm{CH_2OH + CH_3  \rightarrow C_2H_5OH}$ &     0& \\
$\mathrm{C_2H_5 + O   \rightarrow C_2H_5O}$ &     0&  \\
$\mathrm{H      + C_2H_5O  \rightarrow C_2H_5OH}$ &     0& \\
\hskip 2cm{\bf Ethanethiol} &&\\
$\mathrm{C_2H_5 +  HS \rightarrow C_2H_5SH}$ &     0& M\\
$\mathrm{CH_2SH + CH_3  \rightarrow C_2H_5SH}$ &     0& M\\
$\mathrm{S      + C_2H_5  \rightarrow C_2H_5S }$ &     0& M\\
$\mathrm{H      + C_2H_5S  \rightarrow C_2H_5SH}$ &     0& M \\
$\mathrm{CH_3 + CH_3S  \rightarrow C_2H_5SH} $&     0 & \\
\hline
{\bf 1-propanol($\rm {CH_3CH_2CH_2OH}$)}&&\\
$\mathrm{C_2H + H_2O    \rightarrow HCCCHO+H}$ &     0& \\
$\mathrm{O + C_3H_3     \rightarrow HCCCHO+H }$ &     0& \\
$\mathrm{H + HCCCHO    \rightarrow HCCHCHO}$ &     1688& \\
$\mathrm{H + HCCHCHO    \rightarrow H_2CCHCHO}$ &     0& \\
$\mathrm{H + H_2CCHCHO   \rightarrow CH_2CH_2CHO}$ &   2891& \\
$\mathrm{H + CH_2CH_2CHO    \rightarrow CH_3CH_2CHO}$ &     0& \\
$\mathrm{H + CH_3CH_2CHO    \rightarrow CH_3CH_2CH_2O}$ &   2274& \\
$\mathrm{H + CH_3CH_2CH_2O    \rightarrow CH_3CH_2CH_2OH}$ &     0& \\
$\mathrm{C_2H_5 + CH_2OH    \rightarrow CH_3CH_2CH_2OH}$ &     0& \\
\hline
{\bf 1-propanethiol(${\rm CH_3CH_2CH_2SH}$)}&&\\
$\mathrm{C_2H + H_2S    \rightarrow HCCCHS+H}$ &     0& \\
$\mathrm{S + C_3H_3     \rightarrow HCCCHS+H }$ &     0& \\
$\mathrm{H + HCCCHS    \rightarrow HCCHCHS}$ &     2167& \\
$\mathrm{H + HCCHCHS    \rightarrow H_2CCHCHS}$ &     0& \\
$\mathrm{H + H_2CCHCHS    \rightarrow CH_2CH_2CHS}$ &     2659& \\
$\mathrm{H + CH_2CH_2CHS    \rightarrow CH_3CH_2CHS}$ &     & \\
$\mathrm{H + CH_3CH_2CHS    \rightarrow CH_3CH_2CH_2S}$ &     734& \\
$\mathrm{H + CH_3CH_2CH_2S    \rightarrow CH_3CH_2CH_2SH}$ &     0&\\
$\mathrm{C_2H_5 + CH_2SH    \rightarrow CH_3CH_2CH_2SH}$ &     0& \\
\hline
\end{tabular}}}
\hskip 5cm {\scriptsize $^H$ reaction taken from \cite{hase92},}
{\scriptsize $^M$ reaction taken from \cite{mull15}}\\
\end{table}

\subsection{Chemical network}
For the purpose of chemical modeling, we use our large gas-grain chemical network 
\citep{das08b,das13a,das13b}.
Gas-grain interactions are considered to mimic the most realistic scenario of the ISM.
We assume that gas and grains are coupled through accretion and various desorption
mechanisms such as thermal, non-thermal \citep{garr06} and cosmic-ray desorption processes.
Our present gas phase chemical network consist of $6628$ reactions between $684$ gas phase
species and surface chemical network consists of $487$ reactions between $316$ surface species.
We adopt our gas phase chemical network from the UMIST 2006 database \citep{wood07}.
Our gas phase network contains some deuterated reactions as well \citep{das15a,das15b,sahu15}. 
For the grain surface reaction network,
we primarily follow \cite{hase92} and for the ice phase deuterium 
fractionation reactions, we follow \cite{case02,caza10,das16}. Though we have the deuterated
species in our network, we are not considering the deuterium chemistry here for the sake of simplicity. 

Here, we are mainly concentrating on the formation of monohydric alcohols and their thiol analogues. 
These molecules are mainly formed on the dust surface. 
Chemical enrichment of the interstellar grain mantle depends on the binding energies ($E_d$)
and barriers against diffusion ($E_b$) of the adsorbed species. The binding energies of these
species are available from past studies \citep{alle77,tiel87b,hase92,hase93}. But
these binding energies mostly pertain to silicates. Binding energy of the
most important surface species (with ice) which are mostly controlling the
chemical composition of the interstellar grain mantle are available from some
recent studies \citep{cupp07,garr13}. We use these energies in our model. For the rest of
the species for which binding energies were unavailable from these papers,
we keep it the same as the past studies. We use binding energies against diffusion
equal to $0.5 E_d$  \citep{garr13} for our calculations.

\begin{table}
\vbox{\centering
\caption{Peak abundance of ice phase alcohols and their thiols with respect to H nuclei in all forms.}
\vbox{
{\tiny 
\begin{tabular}{|c|c|c|c|c|}
\hline
&\multicolumn{2}{|c|}{\bf Isothermal phase}&\multicolumn{2}{|c|}{\bf Warm-up phase}\\
{\bf Species}&{\bf gas phase}&{\bf ice phase}&{\bf gas phase (temp in K)}&{\bf ice phase (temp in K)}\\
\hline\hline
{\bf Methanol}&$1.75 \times 10^{-9}$&$2.12 \times 10^{-5}$&$6.98 \times 10^{-6}(106.0)$&$1.86 \times 10^{-5}(10.24)$\\
\hline
{\bf Ethanol}&$8.38 \times 10^{-11}$ &$4.66 \times 10^{-11}$ &$1.26 \times 10^{-6}(120.2)$ &$2.00 \times 10^{-6}(71.4)$\\
\hline
{\bf 1-propanol }&$6.45 \times 10^{-20}$ &$1.84 \times 10^{-17}$ &$4.29 \times 10^{-7}(120.6)$ &$4.76 \times 10^{-7}(106.1)$\\
\hline
{\bf Methanethiol}&$1.06 \times 10^{-10}$ &$1.62 \times 10^{-8}$ &$2.16 \times 10^{-8}(107.3)$ &$4.58 \times 10^{-8}(31.2)$\\
\hline
{\bf Ethanethiol}&$3.04 \times 10^{-22}$ &$5.23 \times 10^{-20}$ &$5.45 \times 10^{-9}(120.0)$ &$1.02 \times 10^{-8}(66.2)$\\
\hline
{\bf 1-propanethiol}&$2.94 \times 10^{-27}$ &$6.00 \times 10^{-25}$ &$3.71 \times 10^{-10}(122.7)$ &$3.73 \times 10^{-10}(108.4)$\\
\hline
\end{tabular}}}}
\end{table}

\begin{table}
\vbox{\centering
{\tiny
\caption{Molecular ratio of some species }
\begin{tabular}{|c|c|c|c|c|}
\hline
&{\bf after $1.0 \times 10^6$ years} & {\bf after $1.5 \times 10^6$ years} & {\bf after $2.0 \times 10^6$ years} & {\bf Observed}\\
\hline
&&&&\\
$\frac{\rm CH_3OH}{\rm C_2H_5OH}$&2663&1879&3.24&45$^t$,$78^{m1}$\\
&&&&\\
$\frac{\rm CH_3OH}{\rm CH_3CH_2CH_2OH}$&$2.08 \times 10^9$&3485&7.79&2700$^t$\\
&&&&\\
$\frac{\rm C_2H_5OH}{\rm CH_3CH_2CH_2OH}$&773843&1.85&2.40&60$^t$\\
&&&&\\
$\frac{\rm CH_3SH}{\rm C_2H_5SH}$&$2.61 \times 10^{11}$&708&4.55&$\geq 21^m$,$3.1^{m1}$\\
&&&&\\
$\frac{\rm CH_3OH}{\rm CH_3SH}$&$2.69$&344.3&112.7&$120^m$, $5700^{m1}$\\
&&&&\\
$\frac{\rm C_2H_5OH}{\rm C_2H_5SH}$&$2.62 \times 10^8$&129&157&$125^m$,$225^{m1}$\\
&&&&\\
\hline
\end{tabular}}}
{$^m$\cite{mull15} from observation,}
{$^{m1}$\cite{mull15} from modeling,}
{$^t$\cite{terc15} from observation,}
\end{table}

Our ice phase network contains other reactions mentioned in \cite{hase92}.
In Table 2, we have shown only some grain phase reactions which may lead to the 
formation of these alcohols and their thiol analogues. 
\cite{drua12} shows that chemistry of sulfur may be very different from the chemistry of 
other chemical elements. They considered sulfur polymers ($\rm{S_n}$) and polysulphanes
($\rm{H_2S_n}$) as the potential candidates of the sulfur refractory residue.
Here, we have considered all the sulfur related reactions used in \cite{drua12}.
For the formation of Methanol, we use the pathways proposed 
by \cite{hase92}. 
Methanethiol production is followed by \cite{mull15}. For the ethanol production, we assume
barrier-less addition between $\rm{C_2H_5}$ and OH radical \citep{hase92}, 
$\rm{CH_2OH}$ and $\rm{CH_3}$ radical and hydrogenation reaction with $\rm{C_2H_5O}$. 
For the production of Ethanethiol,
we use the pathways proposed by \cite{mull15}. Reaction references are also
noted in Table 2. Since for the formation of 1-propanol and 1-propanethiol, no pathways
were available, we use some new pathways for the formation of these species in ice phase.
For the formation of the 1-propanol, we have considered two radical-molecular 
ice phase reactions followed by $4$ successive hydrogen addition reactions. Similar sequence is also
considered for the formation of 1-propanethiol. 
In addition, we also have considered the radical radical reaction between $\rm{C_2H_5}$ and
$\rm{CH_2OH}$ for the formation of 1-propanol and radical-radical reaction between $\rm{C_2H_5}$
and $\rm{CH_2SH}$ for the formation of 1-propanethiol.
As like the other radical-molecular
reactions considered in \cite{hase92}, here also, we are assuming the barrier-less nature
of these reactions. Rate coefficients of this type of reactions thus
depend upon the adopted adsorption energies and would process in each encounter. Among the four successive 
hydrogen addition reactions considered here, hydrogen addition reaction in 
second and fourth steps of 1-propanol and 1-propanethiol would be considered as
radical-radical interaction and thus barrierless in nature. 
But the first and third steps of this sequence is the neutral-neutral reaction
which must contain some activation barrier. We have carried out quantum chemical calculation 
to find out suitable transition states for these neutral-neutral reactions. QST2 calculation 
with B3LYP/6-31+G(d) method is employed for this computation and obtained
activation barriers for these neutral-neutral reactions are pointed out in the second column of Table 2.
Though 2-propanol and 2-propanethiol are the structural isomers of 1-propanol and 
1-propanethiol respectively, we are not considering their formation in the present study. For the
destruction of ice phase species, we consider the photo-dissociation reactions by direct interstellar
photons and cosmic ray induced photons.

We do not use any new gas phase formation of these species. 
In our model, gas and grains are continuously
interacting with each other and exchanging their chemical components. 
Surface species could
populate the gas phase by various evaporation mechanism considered here namely; thermal
desorption, cosmic ray induced desorption and reactive  non-thermal desorption (here, we assume
a non-thermal desorption factor to be  $0.01$). For the destruction of gas phase alcohols and 
their corresponding thiols, we use destruction by most abundant ions 
(${\rm {H_3}^+,\ {CH_4}^+, \ C^+, \ HCO^+, \
N^+, \ O^+, {H_3O}^+,\ CH^+,\ {O2}^+,}$  ${\rm H^+, \ He^+, \ {CH_3}^+}$), dissociative
recombination, photo-dissociation and dissociation by cosmic rays.

\subsection{Physical condition}
In order to realistically model the physical parameters, we consider a 
warm-up model \citep{quan16}. Initial phase of this model is the isothermal phase ($T=10$ K) 
followed by a warm-up phase. Both phases have the same constant density ($n_H = 10^4$ cm$^{-3}$) and 
a visual extinction of $10$. Second phase starts with $10$ K and ends at $200$ K. Here, it is 
assumed that the isothermal phase lasts for $10^6$ years and the warm-up phase for another $10^6$ years.
Initial abundances are taken from \cite{drua12} except the sulfur abundance. \cite{drua12} considered
abundance of $S^+$ in its cosmic value  $\sim 1.5 \times 10^{-5}$ \citep{sofi94}. Here, we are assuming much 
reduced $S^+$ abundance ($8.0 \times 10^{-8}$) as used in \cite{leun84}.
Hydrogens are mostly assumed to be in the form of molecular hydrogen. 
These molecular hydrogens were mainly formed on 
the dust surfaces \citep{biha01,chak06a,chak06b} in earlier stages. 
For the ionization of the medium, we assume a cosmic ray ionization rate 
of $1.3 \times 10^{-17}$ s$^{-1}$.

\begin{table}
\scriptsize
\vbox{
\centering{
\caption{vibrational frequencies of 1-propanethiol  and 2-propanethiol in water ice phase
at B3LYP/6-311g++(2df,2pd) method and basis set}
\vspace{1cm}
\begin{tabular}{|c|c|c|c|c|c}
\hline
{\bf Species}&{\bf Peak position} &{\bf Integral absorbance  }&{\bf Band }
&{\bf experimental values}\\
&{\bf in cm$^{-1}$}&{\bf coefficient }&{\bf assignment}
& {\bf wavenumber}\\
&{\bf (in $\mu$m)}& {\bf  in cm molecule$^{-1}$} & &{\bf (in cm$^{-1}$)}\\
\hline
&112.38 ({88.98})&2.30$\times 10^{-19}$&skeletal deformation&\\
&191.24 ({52.29})&3.07$\times 10^{-18}$&SH torsion&\\
&231.45 ({43.20})&8.92$\times 10^{-20}$&CH$_3$ torsion&\\
&243.06 ({41.14})&1.07$\times 10^{-18}$&CH$_3$ torsion&\\
&358.01 ({27.93})&5.51$\times 10^{-20}$&CCC bending&\\
&693.54 ({14.41})&1.69$\times 10^{-18}$&CS stretching&700$^{a}$\\
&733.09 ({13.64})&1.19$\times 10^{-18}$&CH$_2$ rocking&728$^{a}$\\
&805.20 ({12.41})&8.53$\times 10^{-19}$&SH out of plane bending&814$^{a}$\\
&896.45 ({11.15})&1.06$\times 10^{-18}$&CH$_3$ bending/CC stretching &\\
&922.64 ({10.83})&3.01$\times 10^{-19}$&CH$_2$ twisting&\\
&1031.61 ({9.69})&7.72$\times 10^{-20}$&CC stretching&\\
&1098.01 ({9.10})&6.91$\times 10^{-19}$&CH$_2$ rocking&\\
&1128.58 ({8.86})&2.17$\times 10^{-18}$&CC stretching&1105$^{a}$\\
1-propanethiol ($Tg$) &1252.95 ({7.98})&9.61$\times 10^{-19}$&CH$_2$ twisting&1243$^{a}$\\
&1280.52 ({7.80})&3.49$\times 10^{-18}$&CH$_2$ wagging&1300$^{a}$\\
&1322.96 ({7.55})&1.82$\times 10^{-19}$& CH$_2$ twisting&\\
&1365.78 ({7.32})&2.40$\times 10^{-19}$&CH$_2$ wagging&1351$^{a}$\\
&1407.51 ({7.10})&4.15$\times 10^{-19}$&CH$_3$ out of plane bending&1384$^{a}$\\
&1465.93 ({6.82})&5.36$\times 10^{-19}$&CH$_2$ scissoring&1456$^{a}$\\
&1482.97 ({6.74})&1.76$\times 10^{-19}$&CH$_2$ scissoring&\\
&1489.36 ({6.71})&1.71$\times 10^{-18}$&CH$_3$ deformation &\\
&1502.89 ({6.65})&2.11$\times 10^{-18}$&CH$_2$ scissoring&\\
&2666.77 ({3.74})&4.79$\times 10^{-19}$&SH stretching&2598$^{a}$\\
&3025.45 ({3.30})&1.63$\times 10^{-18}$&CH$_3$/CH$_2$ symmetric stretching&2838$^{a}$\\
&3028.23 ({3.30})&4.34$\times 10^{-18}$&CH$_2$ symmetric stretching&2848$^{a}$\\
&3056.01 ({3.27})&1.053$\times 10^{-18}$&CH$_2$ antisymmetric stretching&2945$^{a}$\\
&3056.83 ({3.27})&6.17$\times 10^{-18}$&CH$_2$ symmetric stretching&2960$^{a}$\\
&3085.19 ({3.24})&8.81$\times 10^{-18}$&CH$_2$ antisymmetric stretching&3090$^{a}$\\
&3091.47 ({3.23})&8.41$\times 10^{-18}$&CH$_3$ antisymmetric stretching&\\
&3105.50 ({3.22})&7.11$\times 10^{-18}$&CH$_2$ antisymmetric stretching&3183$^{a}$\\
\hline
\hline
&225.30 ({44.38})&3.70$\times 10^{-18}$&SH torsion&185$^{b}$\\
&230.02 ({43.47})&8.50$\times 10^{-20}$&CH$_3$ torsion&230$^b$\\
&254.64 ({39.27})&4.23$\times 10^{-21}$&CH$_3$ torsion&245$^b$\\
&323.81 ({30.88})&4.81$\times 10^{-19}$& CCS bending&325$^b$\\
&333.83 ({29.95})&4.51$\times 10^{-20}$&CCC bending&\\
&407.72 ({24.52})&5.05$\times 10^{-20}$&CCC bending&410$^b$\\
&594.92 ({16.80})&2.09$\times 10^{-18}$&CS stretching&620$^b$\\
&855.73 ({11.68})&1.84$\times 10^{-18}$&SH out of plane bending&\\
&886.90 ({11.27})&1.11$\times 10^{-19}$&CC stretching&\\
&938.74 ({10.65})&3.70$\times 10^{-20}$& CH$_3$ bending  &\\
&955.04 ({10.47})&1.01$\times 10^{-19}$&CH$_3$ bending &\\
&1101.43 ({9.0})&7.39$\times 10^{-18}$&CH$_3$ bending&\\
2-propanethiol ($Trans$) &1126.78 ({8.87})&3.61$\times 10^{-19}$& CC stretching&\\
&1187.27 ({8.42})&1.36$\times 10^{-18}$& CH$_3$ bending&\\
&1297.52 ({7.70})&4.84$\times 10^{-18}$&CH out of plane  bending&\\
&1336.78 ({7.48})&1.84$\times 10^{-19}$&CH bending  & \\    
&1400.78 ({7.13})&1.39$\times 10^{-18}$&CH$_3$ out of plane bending&\\
&1416.86 ({7.05})&7.64$\times 10^{-19}$&CH$_3$ out of plane bending&\\
&1476.26 ({6.77})&5.41$\times 10^{-21}$&CH$_3$ deformation&\\
&1479.59 ({6.75})&9.35$\times 10^{-19}$&CH$_3$ deformation &\\
&1488.11 ({6.71})&3.01$\times 10^{-18}$&CH$_3$ deformation&\\
&1493.55 ({6.69})&1.64$\times 10^{-18}$&CH$_3$ deformation &\\
&2666.62 ({3.75})&5.06$\times 10^{-19}$& SH stretching&\\    
&3021.65 ({3.30})&4.78$\times 10^{-18}$&CH$_3$ symmetric stretchin &\\
&3027.43 ({3.30})&8.42$\times 10^{-18}$&CH$_3$ symmetric stretching &\\
&3050.92 ({3.27})&7.52$\times 10^{-19}$&CH stretching &\\
&3078.79 ({3.24})&8.99$\times 10^{-20}$&CH$_3$ antisymmetric stretching &\\
&3086.86 ({3.23})&1.50$\times 10^{-17}$&CH$_3$ antisymmetric stretching&\\
&3107.55 ({3.21})&5.08$\times 10^{-18}$&CH$_3$ antisymmetric stretching&\\
&3109.97 ({3.21})&7.21$\times 10^{-18}$&CH$_3$ antisymmetric stretching&\\

\hline
\end{tabular}}}
{$^a$\cite{torg70} and references therein.}
{$^b$\cite{smit68} from experiment}
\end{table}

\subsection{Modeling results}
In Fig. 3, we have shown the time evolution of gas phase (solid curve) and ice phase (dotted curve) 
alcohols and their thiol analogues. Upper panel shows the isothermal phase and lower panel shows 
the warm-up phase. In the isothermal phase, it is clear that ice phase 
methanol, ethanol and methanethiol are efficiently produced. Some portions of these abundant 
ice phase species is readily transfered to the gas phase via various desorption mechanisms. At the
beginning of the warm-up phase, ice phase production of ethanol, ethanethiol, 1-propanol and 1-propanethiol
increases due to the increase in the mobility of the surface species involved in the
reactions. In Table 3, we have pointed out the peak abundances of these alcohols and their
thiol analogues for both the phases. In the warm-up phase, peak abundances
of these species along with the temperatures related to these peak values are also pointed out.

It is fascinating to indicate from Table 3 that among all the species shown in Table 3,
methanol is the only one which is most efficiently produced in the isothermal ($T=10$ K) phase compare
to the warm-up phase. Its peak ice phase abundance in isothermal phase is found to be 
$2.12 \times 10^{-5}$ with respect to
total H nuclei whereas in the warm-up phase, its peak abundance of $1.86 \times 10^{-5}$ is appearing 
around $10.24$ K. In compare to the isothermal phase, abundances of the other ice phase species 
are seemed to be significantly higher in 
the warm-up phase. For example, peak abundance of ice phase methanethiol appears around $31$ K, 
production ethanol and ethanethiol is found to be efficient around $66-71$ K and efficient production
of 1-propanol and 1-propanethiol is found to be around $106-108$ K. Formation of ethanol, 
ethanethiol, 1-propanol and 1-propanethiol at such high temperatures occurs mainly due to the
radical radical reactions. It is essential to point out that adopted adsorption energies 
of some of these key radicals ($\rm {CH_3, \ C_2H_5, \ OH, \ SH, \ CH_2OH, \ CH_2SH}$ are 
$1175$ K, $2110$ K, $2850$ K, $1500$ K, $5080$ K, $5084$ K) available from some earlier studies 
\citep{garr13,cupp07,hase93}.

Since, we are mainly considering the ice phase production of these species, appearance
of the peak gas phase abundance is highly related to their respective adsorption
energies. For example, in case of methanol and methanethiol, we have assumed the adsorption energy $5530$ K
and $5534$ K respectively and from Table 3, the resulting peak gas phase abundances of 
methanol and methanethiol seems around $106-107$ K. For the ethanol, ethanethiol,
propanol and 1-propanethiol much higher adsorption energies are assumed ($6260$ K, $6230$ K, $6260$ K 
and $6260$ K for ethanol, ethanethiol, 1-propanol and 1-propanethiol respectively) which ensures 
the peak gas phase abundance of these species around $120-123$ K.

In Table 4, we have shown molecular ratio (gas phase) of these alcohols and their thiol analogue.
Since, chemical evolution is  highly time dependent phenomenon, 
ratios are shown for various time scales. 
$1.0 \times 10^6$ years corresponds to the end of the isothermal phase, $1.5 \times 10^6$ years 
corresponds to the middle age of the warm-up phase and $2.0 \times 10^6$ years is related to the
end of the warm-up phase.
Gas phase ratio of the observed and other modeling 
results are also shown. Gas phase observational ratios are taken from \cite{terc15}
and \cite{mull15} whereas hot core modeling results is taken from 
\cite{mull15}. It is interesting to note that around the isothermal phase, gas phase
abundance of methanol, methanethiol and ethanol is in the range of $10^{-9}-10^{-11}$ whereas
the gas phase abundances of other species is negligible which yields a much higher molecular ratios
of some species. Beyond $1.0 \times 10^6$ years, mobility of the surface species rapidly increases
and yields significant production of negligible species. At the end of warm-up phase, we are having
a reasonable values of these ratios.

\begin{table}
\scriptsize
\vbox{
\centering{
\caption{Rotational, quartic and sextic centrifugal distortion constants of 1-propanethiol and 2-propanethiol}
\begin{tabular}{|p{1.1 in}|p{0.7in}|p{1.1in}|p{0.85in}|p{0.7in}|p{1.9in}|}
\hline
{\bf Species} & {\bf Rotational constants with equilibrium (e) \&  ground vibrational state (0) geometry} & {\bf Values in MHz with DFT(HF) method}&{\bf Experimentally obtained ground-state values in MHz}&{\bf Distortional constants}
&{\bf Values in KHz with DFT(HF) method} \\
\hline
&A$_e$& 24213.642({24429.75})&&$\Delta$$_J$&0.296911(0.208512)\\  
&B$_e$& 2312.864({ 2337.88})&&$\Delta$$_K$&214.57798(618.4248)\\
&C$_e$& 2222.041({ 2245.40})&&$\Delta$$_J$$_K$&29.811(42.2776)\\
&&&&$\delta$$_1$&-0.0455(-0.039844)\\
&A$_0$&{23239.81}({ 23632.75})&23429.0&$\delta$$_2$&0.4518(9.96033)\\
&B$_0$&{2301.32}({ 2328.06})&2345.29&$\Phi$$_J$&-0.90726$\times 10^{-08}$(-0.2571$\times 10^{-07}$)\\     
1-propanethiol ($Tg$)&C$_0$&{ 2199.16}({ 2226.75})&2250.18&$\Phi$$_K$ & 0.419723$\times 10^{-02}$(0.45931$\times 10^{-01}$)\\   
&&&&$\Phi$$_J$$_K$&0.153284$\times 10^{-04}$(0.48892$\times 10^{-05}$)\\     
&&&&$\Phi$$_K$$_J$ & 0.863327$\times 10^{-03}$( 0.40468$\times 10^{-02}$)\\ 
&&&&$\phi$$_J$ &-0.454229$\times 10^{-07}$(-0.20575$\times 10^{-07}$)\\    
&&&&$\phi$$_K$ &0.194759$\times 10^{-02}$(0.11375$\times 10^{-02}$)\\    
&&&&$\phi$$_J$$_K$&0.557817$\times 10^{-05}$(0.31203$\times 10^{-08}$)\\
\hline
&A$_e$&7886.965({ 7938.21})&&$\Delta$$_J$&1.246( 1.043)\\  
&B$_e$&4341.565({ 4399.57})&&$\Delta$$_K$&6.799(5.473)\\
&C$_e$&3118.889({ 3152.54})&&D$_J$$_K$& 2.184(3.312)\\
&&&&$\delta$$_1$& 3.728(0.265)\\
&A$_0$&{7782.14}({7841.42})&7892.65&$\delta$$_2$&0.3805(1.933)\\
&B$_0$&{ 4306.51}({4366.93})&4414.42&$\Phi$$_J$&0.108001$\times 10^{-07}$( 0.15733$\times 10^{-06}$)\\     
2-propanethiol ($trans$) &C$_0$&{ 3087.77}({3124.21})&3158.03&$\Phi$$_K$ & 0.144633$\times 10^{-04}$(0.54329$\times 10^{-05}$)\\   
&&&&$\Phi$$_J$$_K$&0.562981$\times 10^{-05}$( 0.13656$\times 10^{-05}$)\\     
&&&&$\Phi$$_K$$_J$ & 0.115468$\times 10^{-04}$(0.19811$\times 10^{-05}$)\\   
&&&&$\phi$$_J$ &0.818191$\times 10^{-07}$(0.71495$\times 10^{-07}$)\\    
&&&&$\phi$$_K$ & 0.503435$\times 10^{-05}$(0.10983$\times 10^{-04}$)\\    
&&&&$\phi$$_J$$_K$&0.439650$\times 10^{-05}$(0.12349$\times 10^{-05}$)\\
\hline
\end{tabular}}}
\hskip 5cm $^k$ \citet{kisi10}, $^l$ \citet{grif75}\\
\end{table}

\begin{figure}
\vskip 0.5cm
{\centering
  \includegraphics[height=12cm,width=12cm]{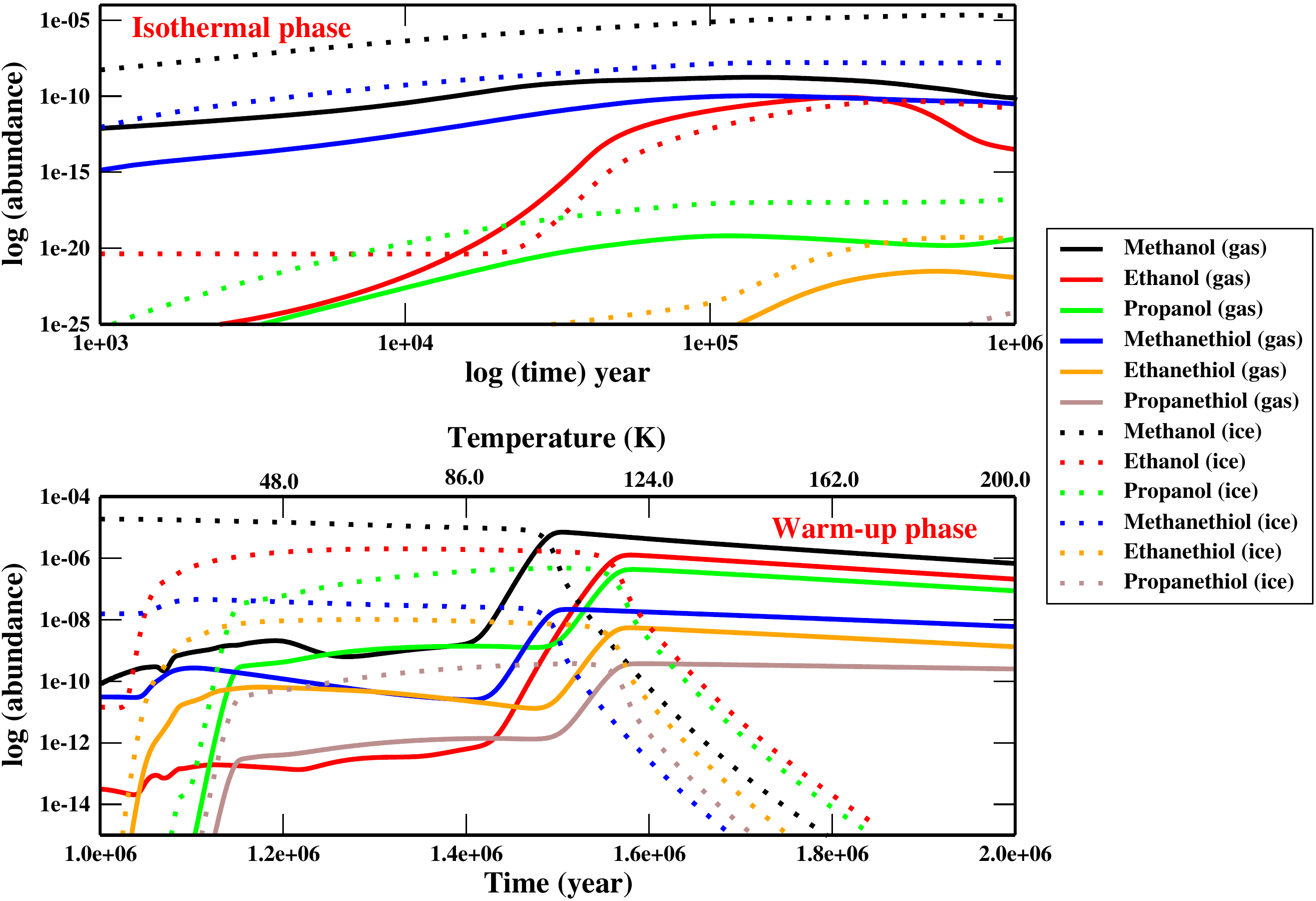}
\caption{\small Time evolution of monohydric alcohols and their thiol analogues in isothermal and
warm-up phase.}}
\vskip 0.5cm
\end{figure}
\section{Spectroscopy}
\subsection{Vibrational Spectroscopy}
Our results suggest that 1-propanethiol would be a probable candidate for the
astronomical detection. Here
we calculate the IR spectrum of 1-propanethiol for the sake of completeness. Moreover,
vibrational spectral information of its one structural isomer, 2-propanethiol is 
also presented. In Table 5, we assigned different modes of vibrations along with 
frequency and intensity of these species. 
Ice phase absorbance is shown in terms of integral absorption coefficient in cm molecule$^{-1}$.
We compare our results with the existing 
experimental results. Gaussian 09 program is used for all these calculations. Water is used as a 
solvent to compute  vibrational spectroscopy using Polarizable Continuum Model (PCM)
with the integral equation formalism variant (IEFPCM) as a default Self-consistent Reaction 
Field (SCRF) method. IEFPCM model is considered to be a convenient one because the second 
derivative of energy with respect to coordinate (bond distance, bond angle) is available for 
this model and also its analytic form is available. For this computations, we use DFT method with 
B3LYP functional and higher order basis set 6-311g++(2df,2pd) \citep{choi08} for better accuracy. 
A comparison between our calculated IR spectrum band  with the existing experimental results 
of 1- proapnethiol and 2-propanethiol \citep{torg70, smit68} are shown in Table 5.  It is
clear from the table that our results are in excellent agreement with the existing
experimental values. Most intense band of 1-propanethiol appears at $3.23$ $\mu$m (3091.47 cm$^{-1}$) and
$3.24$ $\mu$m (3085.19 cm$^{-1}$) due to CH$_3$ and CH$_2$ stretching band respectively 
with the integral absorbance 
coefficient of 8.81$\times 10^{-18}$ and 8.41$\times 10^{-18}$
cm molecule$^{-1}$ respectively. Most intense band of 2-propanethiol belongs to 
3.25 $\mu$m (3078.79 cm$^{-1}$)
which corresponds to the integral absorbance 
coefficient of $1.50 \times 10^{-17}$ cm molecule$^{-1}$. 

Figure 4 shows isotopic variation of vibrational spectra of 1-propanethiol. We show isotopic variation by changing the mass of carbon ($C=12$ and $13$ isotopic mass)  
and sulfur atoms ($S=32$ and $34$ isotopic mass). The result shows that bending mode and 
stretching modes are shifted towards lower wavenumbers. CS stretching for  
${\rm CH_3C{H_2}^{12}C{H_2}^{32}SH}$ mode with 
wavenumber $700.4$ cm$^{-1}$ is shifted to $698.01$ cm$^{-1}$, $\rm{CH_2}$ 
wagging  mode having wavenumber 
$1271.69$ cm$^{-1}$ is shifted to $1265.14$ cm$^{-1}$ and CH$_2$ antisymmetric stretching with wavenumber 
$3111.92$ cm$^{-1}$ is shifted to the wavenumber $3101.18$ cm$^{-1}$ due to change of isotopic mass 
of a carbon atom of CH$_2$ group (CH$_3$CH$_2$$^{13}$CH$_2$$^{32}$SH). 

\begin{table}
\scriptsize
\addtolength{\tabcolsep}{-4pt}
\centering{
\caption{Dipole moments of alcohols and their thiol analogues by using HF/6-31g(d). 
Experimental values are given within the bracket.}
\begin{tabular}{|c|c|c|c|c|}
\hline
&\multicolumn{4}{c|}{\bf Dipole moment components in Debye}\\
\cline{2-5}
{\bf Species}&$\mu$$_a$ &$\mu$$_b$ &$\mu$$_c$ &$\mu$$_T$$_o$$_t$$_a$$_l$ \\
\hline
Methanol &-1.5406 (1.44$^a$) &1.0537 (0.899$^a$) &0.0 &1.8665 (1.69$^a$)\\
Methanethiol &1.4683(1.312$^b$) &1.0152(0.758$^b$) &-0.0001 &1.7851(1.51$^b$)\\	
Ethanol &-0.0541(0.046$^c$) &1.7374(1.438$^c$) &0.0000 &1.78383$^{HF}$,1.53$^{DFT}$(1.441$^c$)\\
Ethanethiol &0.0431(1.06$^d$) &1.8597(1.17$^d$)&0.00 &1.8602 (1.50$^d$)\\
1-propanol ($Gt$) &0.8018,0.574$^x$ (0.32$^{e1}$,0.4914$^{e2}$) &1.0022, 1.086$^x$(1.23$^{e1}$,0.9705$^{e2}$) &1.0743, 0.922$^x$(0.94$^{e1}$,0.9042$^{e2}$) 
&1.6737, 1.53$^x$(1.58$^{e1}$,1.4145$^{e2}$)\\
1-propanethiol ($Tg$) &1.7638&-0.0840 & 0.8186 &1.9463(1.6$^f$)\\
2-propanol ($gauche$) & -1.2070(1.114$^g$)&-0.7023(0.737$^g$) &0.9868(0.8129$^g$) &1.9163(1.56$^g$)\\
2-propanethiol ($trans$)& 0.4034& 1.8685& 0.00& 1.9115 (1.61$^f$)\\
\hline
\end{tabular}}
$^a$\cite{ivas53},
$^b$\cite{tsun89},
$^c$\cite{taka68},
$^d$\cite{schm75},
$^e$\cite{abdu70},
$^f$\cite{lide01},
$^g$\cite{hiro79a},
$^x$calculation at MP2/cc-pVTZ level.
\vskip 0.5cm
\end{table}

\begin{figure}
\centering
\includegraphics[width=14cm, height=11cm]{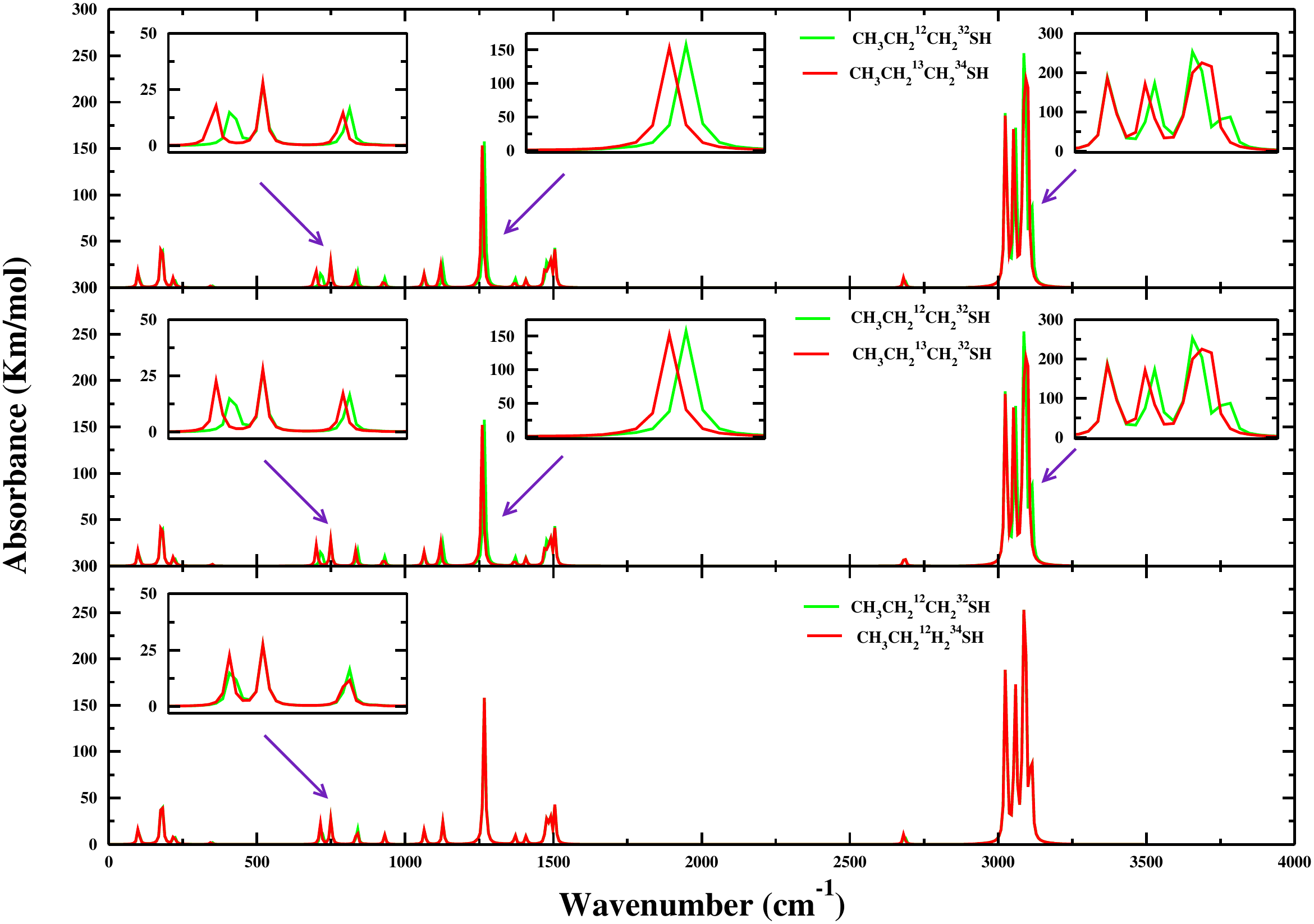}
\caption{Isotopic variation of infrared  spectra of 1-propanethiol.}
\end{figure}
 \begin{table}
\scriptsize
\centering{
\caption{Expected intensity ratio by assuming the same column density and rotational temperature.}
\begin{tabular}{|c|c|}
\hline
&ratio\\
\hline
$\rm \frac{Ethanol}{Methanol}$&0.435\\
&\\
$\rm \frac{1-Propanol}{Methanol}$&0.181\\
&\\
$\rm \frac{2-Propanol}{Methanol}$&0.179\\
&\\
$\rm \frac{Methanethiol}{Methanol}$&1.163\\
&\\
$\rm \frac{Ethanethiol}{Methanol}$&0.249\\
&\\
$\rm \frac{1-Propanethiol}{Methanol}$&0.116\\
&\\
$\rm \frac{2-Propanethiol}{Methanol}$&0.111\\
\hline
\end{tabular}}
\vskip 0.5cm
\end{table}

\subsection{Rotational Spectroscopy}
Till date, most of the species are observed in the interstellar medium or circumstellar shells by their
rotational transitions. 
\cite{chak15,maju14a,maju14b,maju13,maju12} pointed out the need for theoretical
calculations for firm identification of some unknown species in the ISM.
Species which have permanent dipole moments show their rotational
transitions. Here, we compute various rotational parameters (for equilibrium structure
as well as ground vibrational state) for 
1-propanethiol and 2-propanethiol.
Here, we have employed B3LYP/aug-cc-pVTZ and HF/cc-pVTZ method in 
Gaussian 09 program.
Aug prefix basis set is used here to mean that the basis set is augmented with diffusion function and 
cc-pVTZ is Dunning correlation consistent basis sets \citep{kend92} having triple zeta function. This
basis set has its redundant functions removed and is rotated \citep{davi96} in 
order to increase computational efficiency. Accuracy depends on the choice of the method 
and basis sets used. Anharmonic  vibrational-rotational coupling analysis is computed using 
the second order (numerical differentiation) perturbative anharmonic analysis. Quartic 
rotation-vibration  coupling  is included in rotational parameters calculations. Calculated 
rotational and distortional constants are shown in Table 6 to compare with some existing results.
It is to be noted that the existing experimental results which are pointed out in Table 6 were
for the ground vibrational state.

Various components of dipole moments are computed for all the alcohols and their thiols considered
in this study. In Table 7, we compare our calculated dipole moment components with the 
existing theoretical or experimental results. Previous studies found that calculations at the
HF level would predict dipole moment components close to the experimental values. Thus, we
use HF/6-31g(d) level of theory for this computation. It is expected that these complex molecules could be detected in hot core regions. 
\cite{char95} pointed out that for an optically thin emission, an idea about the antenna 
temperature could be made by calculating the intensity of a given transition. This intensity
is proportional to $\frac{{\mu}^2}{Q(T_{rot})}$, where $\mu$ is the 
electric dipole moment and $Q(T_{rot})$ is the partition function at rotational temperature $T_{rot}$. 
In Table 8, we compare the intensities for all the species with respect to
methanol. For the computation of $Q(T_{rot})$, we use $\sqrt{T^3/(ABC)}$. 
Rotational constants of these species are taken from earlier studies 
\citep{taka68,ohas77,hiro79,sast86,luci89,kisi10,mull15,grif75}. Here, we assume that
all these species bear the same column density and rotational temperature. 
Since we are aiming to study these molecules around hot
core regions, we are using $T=180$K for this calculation. All these ratios are shown in Table 8. 
Very nice correlation is seen  as we going to higher order alcohols/thiols. 
The spectral intensities , along with the frequencies
for rotational transitions of 1-propanethiol and 2-propanethiol in the sub-millimeter regime are predicted by
using quantum chemical calculations followed by the SPCAT program \citep{pick91}.
For this calculations, we use the experimentally obtained constants from Table 6 and use
experimentally obtained dipole moments from Table 7. We prepare this catalog files in JPL format and
this files are given as supplementary materials with this article.

\section{Conclusions}
In this paper, we study the formation of monohydric alcohols and their thiols. Major highlights
of our work are as follows.\\

$\bullet$ In between various conformational isomers, it is essential to find out
the most stable conformer which might be a viable candidate for astronomical detections. Here, we
carried out potential energy surface scan to find out the most stable isomer of the monohydric
alcohols and their thiol analogues. Among the alcohols, methanol, ethanol and 1-propanol
have been claimed to be detected in the ISM whereas in thiols, methanethiol and ethanethiol were 
claimed to be detected in hot core regions. In between alcohols, 2-propanol and
in between thiols, 1-propanethiol and 2-propanethiol are yet to be detected in
any sources. Our calculations find that 
$gauche$, $Tg$ and $trans$ conformer is the most stable isomer for 
2-propanol, 1-propanethiol and 2-propanethiol respectively.\\

$\bullet$ Reaction pathways in forming all stable isomers of monohydric alcohols and their 
thiols are prepared to study the chemical evolution.\\

$\bullet$ {Our study reveals that around the warmer region ({$T> 120$ K}), 1-propanethiol would be a 
viable candidate for astronomical detection in the gas phase.}\\

$\bullet$ Since 1-propanethiol is yet to be detected in space, we carried out quantum chemical
calculation to study various spectral aspects (in IR and sub-mm) of this species.
Band assignments were done for its various modes of vibration. 
Changes of absorbance spectra due to the isotopic effects were also pointed out.
Moreover, we find out rotational and distortional constants of this species and compare
with existing experimental results. Experimentally obtained constants and our calculated dipole
moment components are further utilized to predict various probable transitions which should be
useful for the future detection of this species in the ISM.

\section{Acknowledgement} 
PG is grateful to DST (Grant No. SB/S2/HEP-021/2013) for the partial financial support.
AD and SKC want to acknowledge ISRO respond project (Grant No. ISRO/RES/2/402/16-17).
EEE acknowledges a research fellowship from the Indian Institute of Science, Bangalore. Amaresh Das
acknowledges the partial support of Inidian Centre for Space Physics.

\end{document}